# Interfacial ferroelectricity in marginally twisted 2D semiconductors

Astrid Weston[1,2], Eli G Castanon[3], Vladimir Enaldiev[1,2,4], Fabio Ferreira[1,2], Shubhadeep Bhattacharjee[1,2], Shuigang Xu[1,2], Héctor Corte-León[3], Zefei Wu[1,2], Nickolas Clark[2,5], Alex Summerfield[1,2], Teruo Hashimoto[2,5], Yunze Gao[1,2], Wendong Wang[1,2], Matthew Hamer[1,2], Harriet Read[1,2], Laura Fumagalli[1,2], Andrey V Kretinin[1,2,5], Sarah J. Haigh[2,5], Olga Kazakova[3], A. K. Geim[1,2], Vladimir I. Fal'ko[1,2,6 *], and Roman Gorbachev[1,2,6 *]

**1** Department of Physics and Astronomy, University of Manchester, Oxford Road, Manchester, M13 9PL, UK
**2** National Graphene Institute, University of Manchester, Oxford Road, Manchester, M13 9PL, UK
**3** National Physical Laboratory, Hampton Road, Teddington, TW110LW, UK
**4** Kotel'nikov Institute of Radio-engineering and Electronics, Russian Academy of Sciences, 11-7 Mokhovaya St, Moscow, 125009 Russia
**5** Department of Materials, University of Manchester, Oxford Road, Manchester, M13 9PL, UK
**6** Henry Royce Institute for Advanced Materials, University of Manchester, Oxford Road, Manchester, M13 9PL, UK

E-mail: vladimir.falko@manchester.ac.uk, roman@manchester.ac.uk.
*First three authors equally contributed.*

## Abstract

Twisted heterostructures of two-dimensional crystals offer almost unlimited scope for the design of novel metamaterials. Here we demonstrate a room-temperature ferroelectric semiconductor that is assembled using mono- or few- layer $MoS_2$. These van der Waals heterostructures feature broken inversion symmetry, which, together with the asymmetry of atomic arrangement at the interface of two 2D crystals, enables ferroelectric domains with alternating out-of-plane polarisation arranged into a twist-controlled network. The latter can be moved by applying out-of-plane electrical fields, as visualized in situ using channelling contrast electron microscopy. The interfacial charge transfer for the observed ferroelectric domains is quantified using Kelvin probe force microscopy and agrees well with theoretical calculations. The movement of domain walls and their bending rigidity also agrees well with our modelling results. Furthermore, we demonstrate proof-of-principle field-effect transistors, where the channel resistance exhibits a pronounced hysteresis governed by pinning of ferroelectric domain walls. Our results show a potential venue towards room temperature electronic and optoelectronic semiconductor devices with built-in ferroelectric memory functions.

Keywords: 2D materials, twistronics, memristors, heterostructures, semiconductors

**Introduction**

Ferroelectrics are crystals that feature intrinsic charge polarisation with two or more preferred stable directions of the polarisation vector, determined by the lattice symmetry. Switching between those stable polarisation states can be controlled by application of an external electric field; allowing various applications including non-volatile memory, microwave devices, transistors, and sensors[1,2]. Of particular interest are ferroelectric semiconductors that would potentially allow field-effect transistors with additional functionality, e.g. storing information. However, it has proven challenging to find suitable materials that will both remain ferroelectric at room temperature and can be manufactured as thin films required by the microelectronics industry[3]. The latter requirement is a severe challenge for traditional oxides due to the adverse effects of interface quality, which limit the realistically achievable homogenous ferroelectric layer thicknesses[1]. An alternative is to use layered crystals that can be cleaved or grown in ultrathin form while retaining surface quality. Only few such materials have been experimentally demonstrated so far, including in-plane ferroelectric SnTe[4], out-of-plane $CuInP_2S_6$[5] and both types of ferroelectricity in different phases of $In_2Se_3$[6,7]. Up to now, out-of-plane switchable ferroelectricity at room temperature was achieved only in films thicker than 3 nm[8].

Recently a new trend in creating truly two-dimensional (2D) ferroelectrics has emerged, which exploits interfacial charge transfer in stacked heterostructures of 2D materials. This possibility has been recently demonstrated in marginally twisted wide band gap insulator, hexagonal boron nitride [9–11] and in semi-metallic $WTe_2$ [12] with the ability to switch the domain type achieved by sliding atomic planes along the interface. Here, we report an observation of robust room temperature ferroelectricity in marginally twisted semiconducting bilayers of the transition metal dichalcogenide (TMD), $MoS_2$. While ferroelectricity in TMD bilayers with parallel stacking of unit cells has been predicted theoretically[13] and layer-polarised electronic band-edge states have been recently observed in electron tunnelling[14] and optical properties[15], the net charge transfer and, therefore, steady-state electrical polarisation has not been shown experimentally. Here we use a combination of electron channelling microscopy, Kelvin-probe force microscopy and electron transport measurements to explore ferroelectric domain networks in $MoS_2$ bilayers and their dependence on the interlayer twist angle. We demonstrate that such domains can be switched by either mechanical shear force, or by an external out-of-plane electric field, focusing on how the switching behaviour evolves with lateral domain size. We also find that domain switching produces a strong a response in the lateral electronic properties of the twisted $MoS_2$ layers. Together with the extraordinary optical properties of TMDs[16], our work offers a promising avenue towards designing novel devices where both memory effect and optoelectronic functionality can be achieved within a single interface in the ultimate 2D limit.

**Main text:**

Layer-by-layer assembly of van der Waals heterostructures from various 2D crystals has matured into a sophisticated experimental field where not only ultra-sharp and ultra-clean interface quality[17] is achieved but also the rotation between adjacent monolayers, or twist, can be controlled with high precision. Introducing twist gives rise to moiré superlattices where periodic variation of a local atomic registry can lead to profound changes in the electronic properties of the resulting system. For small twist angles ($\theta<2°$) atomic lattices in TMDs often undergo substantial reconstruction due to the energy gain from the preferential stacking domains balancing the cost of the resulting lattice strain[18,19]. In homobilayers, such as $MoS_2/MoS_2$ studied here, the period $\ell = a/\theta$ of the domain network is much longer than the lateral lattice constant, $a$, of $MoS_2$ and determined by the twist angle, $\theta$. If the two $MoS_2$ layers are oriented 'parallel', they form a bilayer of bulk 3R polytype when $\theta = 0$, and 3R commensurate domains formed at small twist angles. Such domains are separated by domain walls (DW) that form a triangular network and have been identified as partial dislocations (PD) in earlier transmission electron microscopy studies[14].

To investigate the possible presence of ferroelectricity we have assembled bilayers of $MoS_2$ using the tear and stamp technique[20] aiming at a zero global misalignment angle. While ideally this could result in a 3R bilayer, small random deformations inflicted by the transfer process[21] lead to a gradual variation of the misalignment angle $|\theta| < 0.1°$. To visualize the resulting domain structure, we adopt back-scattered electron channelling contrast imaging (BSECCI), previously used for bulk materials[22]. We find that this techniques provides a high contrast even for twisted bilayers encapsulated under hexagonal boron nitride (hBN) crystals that were several nanometres thick, similar to the recently used channelling-modulated secondary electron imaging[23]. An example of BSECCI in Fig. 1a shows triangular domains of varying sizes (from 100 nm to >1μm), which enables us to establish and quantify the dependence of domain behaviour on their lateral dimensions. The stacking order of the two domain types (seen as regions of dark and light contrast in Fig. 1a) is illustrated schematically in Fig. 1b and denoted as $Mo^tS^b$ ($S^tMo^b$) corresponding to the vertical alignment of molybdenum atomic positions in the top layer with sulphur positions in the bottom (and *vice versa*). As $Mo^tS^b$ can be seen as a mirror image of $S^tMo^b$ stacking, they feature equal adhesion energies and therefore are expected to occupy similar areas of the sample with no clear preference, in agreement with our observations.

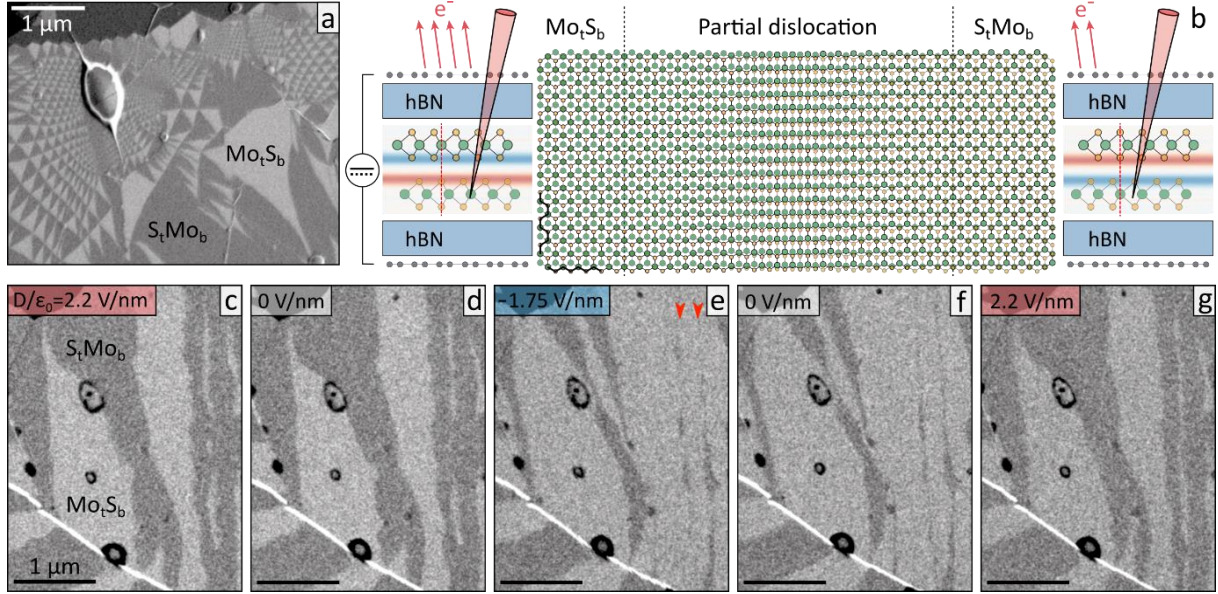

**Figure 1: Ferroelectric domains in marginally twisted MoS₂. (a)** Example of BSECCI acquired on twisted bilayer MoS₂ placed onto a graphite substrate. Light and dark domain contrast corresponds to the two dominant stacking orders referred as Mo$^t$S$^b$ and S$^t$Mo$^b$ respectively. **(b)** Centre - a schematic demonstrating the transition from Mo$^t$S$^b$ to S$^t$Mo$^b$ with perfectly stacked bilayer regions separated by a partial dislocation. Side panels show cross-sectional alignment of the MoS₂ monolayers along the armchair direction assembled within the double gated device structure. **(c-g)** Domain switching visualised by BSECCI under transverse electric field applied *in situ*. Large domains mostly retain their shape when the field is removed and practically disappear when the field is inverted; the arrows in (e) indicate partial dislocations colliding when neighbouring domains of the same orientation try to merge. Micrographs are presented in chronological order. White oval feature in (a) and black ring features in (c-g) are where the intralayer contamination has segregated to form a bubble.

In contrast to TMD bilayers with 2H stacking which possesses both $C_3$ rotational and inversion symmetry, the lattice of 3R bilayers is only $C_3$-symmetric, having neither an inversion centre nor a mirror reflection plane. This asymmetry allows for a steady-state electrical polarisation, which has been shown theoretically[13] to result from an interlayer charge transfer due to asymmetric hybridisation between the conduction band states in one (e.g., top) layer and the valence band states in the other (e.g., bottom) layer. Charge density transferred between the layers (red for positive and blue for negative charges), computed using density functional theory (implemented in the *Quantum Espresso* code[24]) and averaged over the MoS₂ unit-cell area, is shown in the side panels of Fig. 1b. This analysis indicates that the resulting double layer of charge resides on the inner sulphur sublayers (see Fig. 1b), generating an areal density, $P = \pm 3.8 \times 10^{-3}\ e/\text{nm}$, for the out-of-plane electric dipole moment in Mo$^t$S$^b$/S$^t$Mo$^b$ domains. This also agrees with the estimation, $P = \epsilon_0 \Delta V^{FE}$ obtained using the DFT-calculated potential of the stacking-dependent double-layer, $\Delta V^{FE}$ ($|\Delta V^{FE}| = 63$mV for 3R stacking), described in Supplementary Section S7. Coupling to this electrical polarisation with an applied electric field favours - depending on the direction of the external field - either Mo$^t$S$^b$ or S$^t$Mo$^b$ stacking, providing means to modify the domain structure.

To observe this ferroelectricity experimentally we encapsulated twisted MoS$_2$ bilayers in hBN and used graphene on both sides to enable electrostatic gating. Electrical contacts to both graphene layers have been made using stencil mask lithography to minimise contamination (see Supplementary Section 1 for further details). We have found that domains can be clearly seen in BSECCI images, despite overlay of the multilayer hBN and graphene. We consider first the effect of applied field on a double gated bilayer region containing elongated stripe domains as shown in Fig. 1c-g. These images show that the domain structure strongly varies on the application and reversal of the out-of-plane electric displacement, $D = \epsilon_0 V \epsilon_r / h$ (where $\epsilon_r = 3.5$ is the relative permittivity of hBN[25,26] and $h$ is the total thickness between the graphene gates in nm), achieved by varying gate voltages, $V$. The domain configuration, prepared by applying $\epsilon_0^{-1} D = 2.2$ V/nm, stays the same upon removing the gate voltage ($D = 0$) (Fig. 1c and 1d). Then, the application of $\epsilon_0^{-1} D = -1.75$ V/nm gradually expands the area of the lighter contrast domain type at the expense of the darker contrast domains. Again, if the gate voltage is swept back, the domain structure remains the same up to $D = 0$ and through a small interval of positive voltages, but then it returns to almost identical configuration as was observed at the beginning of the hysteresis cycle (cf. Fig. 1c and 1g). A more detailed study of the domain evolution upon sweeping $D$ is shown in Supplementary Fig. S2. Importantly, although the 'darker contrast' domains in in Fig. 1e appear to be 'squeezed' to almost unnoticeable width (marked by red arrows), they still remain visible as thin line defects within the expanded light contrast domains. Upon reversal of the field, these lines serve as precursors for growing domains of opposite polarisation (dark contrast in Fig. 1g). This behaviour indicates that a pair of partial dislocations at Mo$^t$S$^b$/S$^t$Mo$^b$ and S$^t$Mo$^b$/Mo$^t$S$^b$ boundaries combines into a topologically protected defect, a perfect screw dislocation (PSD) between two domains of the same orientation (barely visible at the spatial resolution of Fig. 1c-g; see further). We also note that some domains change relatively little with gate voltage (see, e.g., the bottom-left domain in Figs. 1c-g), which can be attributed to DW pinning by structural imperfections (sample edges, hydrocarbon bubbles, *etc*.).

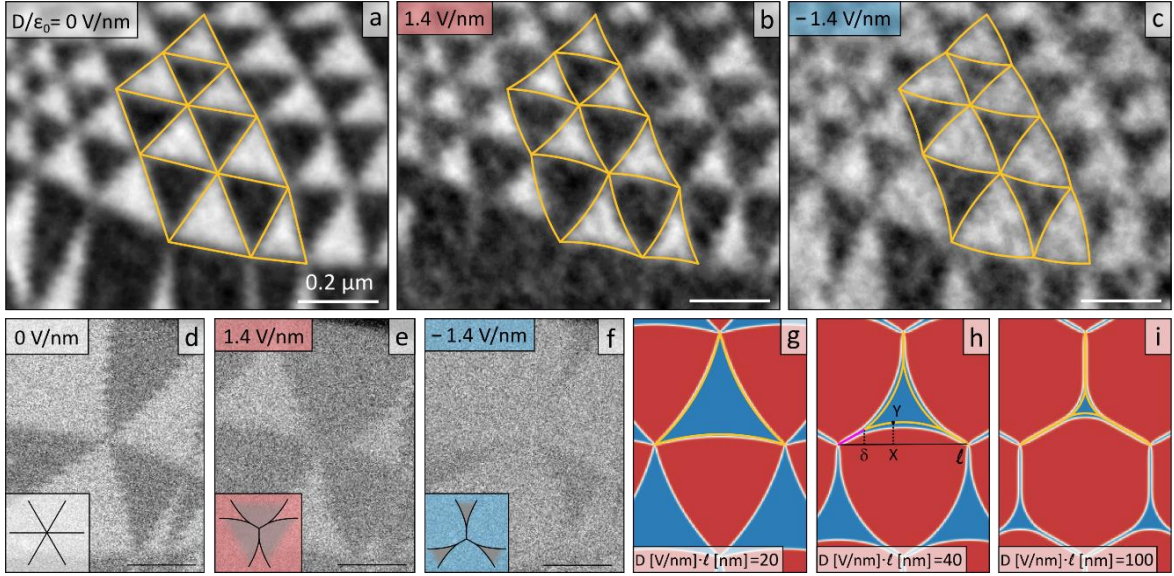

**Figure 2: Domain evolution in double-gated marginally twisted MoS₂ bilayers.** A triangular network of small domains in **(a-c)** undergoes expansion/contraction as a function of applied electric field overlaid with the analytical model Eqs. (2,3), yellow lines. For larger domains in **(d-f)**, the partial dislocations that constitute the domain walls merge near the nodes and the energetically disadvantaged domain collapses locally into a PSD. Micrographs are presented in their chronological order and the contrast is seen to deteriorate due to beam-induced surface contamination. Scale bars are 200 nm. **(g-i)** Polarisation maps computed using mesoscale relaxation of the bilayer lattice (see Supplementary Section S9) and compared with the analytical model (yellow curves) of the scaled domain evolution given by Eqs. (2) and (3).

A strikingly different behaviour is observed in the areas hosting triangular domain networks on application of electric field, Fig. 2a-c. Similar to Fig. 1c-g, positive field favours the darker contrast domains and negative field favours the lighter contrast. However, in these triangular networks the nodes, where three domain walls intersect, remain fixed for all electric displacements. The domains expand (contract) by concave (convex) curvature of the domain walls, with the degree of curvature changing continuously with applied electric field. Rounding of the walls starts as soon as the electric field is applied, without a discernible threshold in electric displacement $D$, and the effect is more pronounced for long domain walls. Characterising domain walls by the minimum distance between the nodes (length, $\ell$, illustrated schematically Fig. 2h) we find that for relatively large domains ($\ell \sim 400$ nm) the walls can be seen to merge (Figs. 2d-f). This begins where the partial dislocations are closest, near the nodes, and causes the triangular domains to shrink to less than 50% of the original size, leaving a perfect screw dislocation (PSD) that connects the smaller triangular domain to the three surrounding nodes. As the domain walls consist of two partial dislocations with Burgers vectors $\frac{a}{\sqrt{3}}(1,0)$ and $\frac{a}{\sqrt{3}}(\frac{1}{2},\frac{\sqrt{3}}{2})$, line defects observed in Fig 1e,f and 2e,f can be assigned to a perfect screw dislocations (PSD) with Burgers vector $a(\frac{\sqrt{3}}{2},\frac{1}{2})$.

To better understand the evolution of triangular domain network we now outline an analytical description of the domain wall behaviour. We describe the domain wall bending as a transverse displacement, $y(0 < x < \ell)$, of a single DW segment from a straight line $[y(x) = 0]$ connecting two nodes. We analyse the variation of energy per supercell of an ideally periodic network (like in Fig. 2g-i), caused by the application of an out-of-plane electric field,

$$\mathcal{E}_\ell[y(x)] \approx 3\int_\delta^{\ell-\delta} \left[\left(\tfrac{1}{2}\bar{w} + \tilde{w}\right)y'^2 - 2\frac{DP}{\epsilon_0 \chi}y\right]dx + 6\left[\frac{u}{\sqrt{3}} - \bar{w} - \frac{\Omega}{\sqrt{3}}\delta\right]\delta \,. \quad (1)$$

The first term in the integral part of $\mathcal{E}_\ell$ is due to elongation of the partial dislocations accounted as $\sqrt{1+y'^2} - 1 \approx \tfrac{1}{2}y'^2$, as well as their energy dependence, $\bar{w} + \tilde{w}\sin^2\phi \approx \bar{w} + \tilde{w}y'^2$, on the deviation angle, $\phi = \arctan y' \approx y' \equiv \frac{dy}{dx}$, of the DW axis from the closest armchair direction. In this parametrisation, we can use the earlier-computed[27] values of $\bar{w} = 0.94$ eV/nm for the energy density of a partial dislocation aligned along the armchair direction and $\tilde{w} = 0.62$ eV/nm for its orientation-dependent part. This approximation assumes that $y'^2 \ll 1$, which is justified by further considerations in Supplementary Section S8. The second term in the integral accounts for the energy gain from the redistribution of domains area, promoted by the displacement field $D$. Here, we describe the coupling between the external field and the ferroelectric polarisation density, $P$, using an effective dielectric screening parameter, $\chi$, which we later estimate based on the comparison with the experimentally observed evolution of domains with various sizes, $\ell$.

The remaining terms in Eq. (1) account for the possible merger of two partial dislocations into a perfect screw dislocation. The PSDs will separate two equivalently polarised domains near the network nodes and will be aligned along the MoS$_2$ zigzag directions (at ±30° from the armchair directions, which are also the partial dislocation orientations in the unperturbed domain network). Formation of a PSD requires a sufficiently high displacement field characterised by a threshold value, $D > D_*(\ell) \propto \ell^{-1}$. The PSDs, are characterised[27] by energy density, $u$ = 2.19 eV/nm, and, when projected onto the intervals $0 < x < \delta$ and $\ell - \delta < x < \ell$, set boundary conditions for the remaining PD segment, as $y(\delta) = y(\ell - \delta) = \delta/\sqrt{3}$.

For $D < D_*(\ell)$, the border between S$^t$Mo$^b$ and Mo$^t$S$^b$ domains is obtained by the minimization of energy in Eq. (1) and has a parabolic shape:

$$y(x) = \frac{DP/\chi\epsilon_0}{\bar{w} + 2\tilde{w}}(\ell - x)x \,. \quad (2)$$

By fitting the experimentally observed evolution of domain shapes for various $\ell$ in Fig. 2a-c with Eq. (2), we estimate that the screening parameter is $\chi \approx 1.5$. The latter value enables us to establish (see in Supplementary Section S8) the domain-length-scale dependent threshold,

$$\epsilon_0^{-1}D_*(\ell) \approx \frac{400\,V}{\ell}, \quad (3)$$

at which pairs of PDs start merging near the network nodes into more energetically favourable PSDs. The merger occurs when the external electric field drives the PD in the neighbouring domains to touch each other (for details, see Supplementary Section S8). For domains with $\ell \sim 400$ nm, this critical regime can be reached within a realistic range of electric fields, as observed experimentally in Figs. 2d-f. Theoretically, we estimate that, for $D > D_*(\ell)$, near each node the PSD segments grow in length, as $\frac{\ell}{\sqrt{3}}\frac{D-D_*}{D}$, whereas shorter PD segments retain an approximately parabolic shape (for comparison with the exact numerical solution, see in Fig. S11c). Figures 2g-i illustrate such an evolution, with the analytical results of Eq. (2) and (3) laid over a polarisation map of a domain structure computed using mesoscale lattice relation (see Supplementary Section S9).

Having established the inverted ferroelectric polarisation inside $Mo^tS^b$ and $Mo^bS^t$ domains, we have also measured the potential, $\Delta V$, created at the bilayer interface, which was predicted to be $\Delta V \approx 65 \pm 5$ mV for all semiconducting TMDs[13,28]. To this end, we employ two-pass phase-modulated Kelvin-probe force microscopy (PM-KPFM) performed in high vacuum. First, the sample topography acquired in the tapping mode with the probe grounded. Then, the probe is lifted at a constant height and scanned following the sample's topography, while an AC voltage, $V_{AC}$, is applied to the probe to create the oscillating electrostatic force which is further minimised by applying a DC voltage. This DC voltage is equal to the contact potential difference between the tip and the sample and allows us to extract the local work function of the surface for a known tip material. For this study, the TMD bilayer was deposited onto a thick hBN on an oxidised silicon wafer, and was contacted using an overlapping graphene layer, followed by a stencil mask deposition of Cr/Au contacts (see Supplementary Section 3).

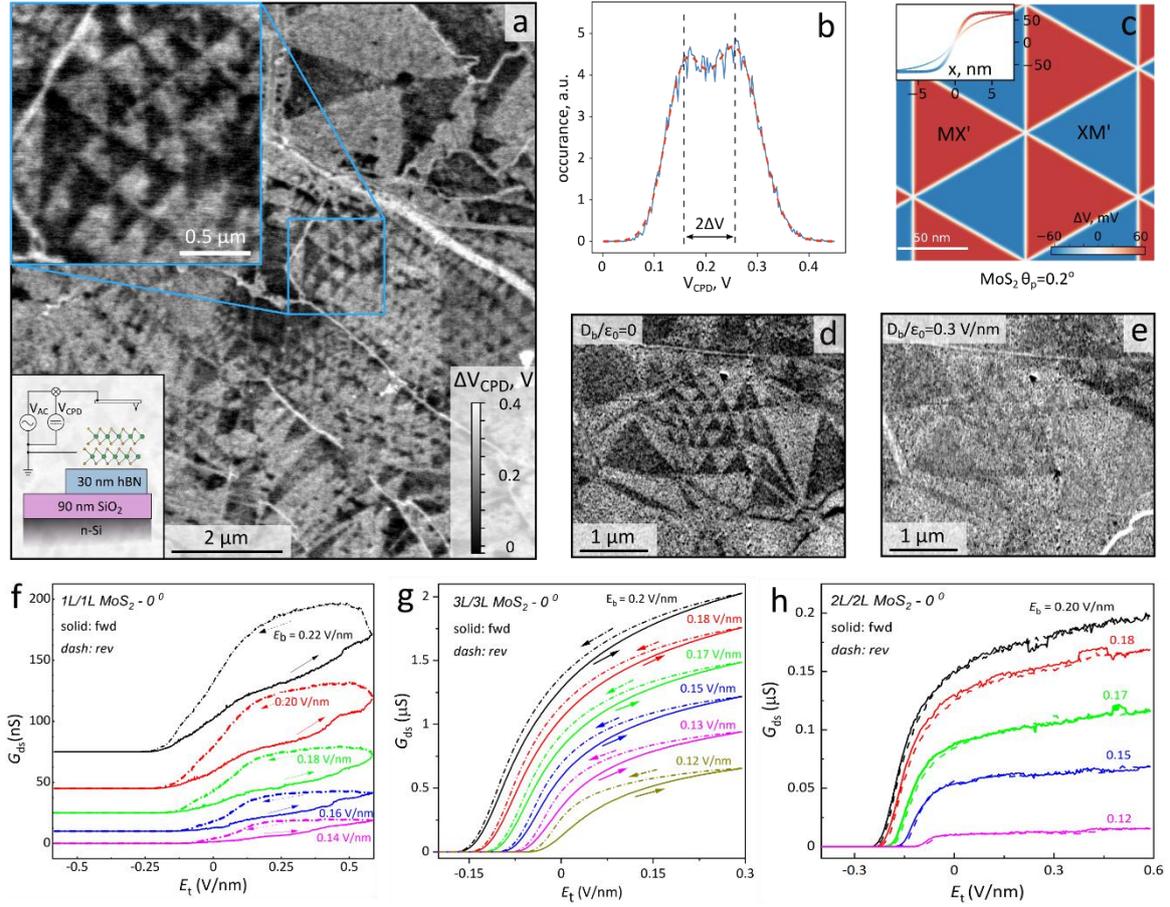

**Figure 3: Electronic properties of ferroelectric domains in MoS₂. (a)** Two-pass PM-KPFM map of the surface potential acquired with $V_{AC}$= 5.75V, and a time-average probe-sample distance of 37 nm during the second pass. The blue boxed inset magnifies the area used to extract numerical values of $2\Delta V$. The lower left inset shows the schematic of PM-KPFM measurement set-up. **(b)** Typical histogram analysis used to extract surface potential difference. In this case, the data from the blue boxed area in (a) was used; for details and analysis of other areas see Supplementary Section 4. **(c)** Calculated surface potential distribution for experimentally relevant domain sizes. Inset shows the potential drop across the domain walls where narrow and bold curves take into account and ignore piezoelectric charges, respectively. **(d)** PM-KPFM surface potential map of a sample with no gate voltage and **(e)** with back gate voltage applied indicating disappearance of the potential variation when free electrons are introduced. **(f-g)** Hysteretic behaviour of electrical conductivity $G_{sd}$ of our artificially made ferroelectric semiconductors as a function of top-gate displacement field ($D_t$) for different back-gate displacement fields ($D_b$). The shown curves are for 1L MoS₂ on top of 1L MoS₂ at 350 K **(f)** and 3L/3L MoS₂ at room temperature **(g)** twisted by 0° to achieve the 3R interface. We used the top gate for recording hysteresis because it covers only the twisted region whereas the bottom gate influences a much larger area, including contact regions. **(h)** similar measurements on a reference 2L/2L sample twisted by 0° which produces 2H interface and displays much smaller hysteresis with the opposite sign (room temperature).

The resulting map of the surface potential acquired on a marginally twisted bilayer MoS₂ is shown in Fig. 3a, where the two types of domain (Mo$^t$S$^b$ and Mo$^b$S$^t$) display a clear contrast difference. This difference, measured with a conductive scanning probe several nanometres above the sample's surface, indicates the presence of a transverse electric field built-in to the domains polarised in the opposite directions for Mo$^t$S$^b$ and Mo$^b$S$^t$ regions, respectively. The gained potential gained due to such fields with respect to the metallic n-doped Si back gate, $\pm\Delta V$, is measured experimentally as a $2\Delta V$ jump when a boundary between two domains is crossed. Having optimised KPFM measurements against lift height and $V_{AC}$ (aiming at low AC driving voltages), we build a histogram of the local potential, see Fig. 3b. Using double Gaussian fitting of the two major peaks (corresponding to the dark and light contrast domains) reveals $2\Delta V = 100 \pm 20$ mV, in close quantitative agreement with the theoretically predicted[13] value, $2\Delta V = 126$ mV, and the potential map in Fig. 3c. The measured $\Delta V$ was found to be reproducible for various locations on the sample and also consistent with the results obtained for another sample placed on graphite substrate (Supplementary Section 5). Upon applying a back gate voltage to the sample, we observed that domain pattern gradually vanishes, as expected due to screening of the ferroelectric field by the free electrons. For electron density ~$10^{13}$cm$^{-2}$ the pattern becomes barely visible, cf. Fig.3d and Fig3f, however the domain shape changes very little since the electric field inside the bilayer in the case of one-sided gating is substantially screened by the bottom MoS$_2$ layer. Note that the piezo-charges reported earlier[14] could partially compensate for the ferroelectric charge transfer. However, piezoelectricity has little impact on the potential in the middle of large domains, as it is caused by strain localised within several nanometres of the domain walls (cf. narrow and bold curves in the inset of Fig. 2c), which is below our lateral resolution.

For a 2D ferroelectric semiconductor, the potential difference between the opposite polarisation of domains should translate into gate-controlled doping and, therefore, electron transport through such ferroelectric field-effect transistor (FFET) devices should depend on the ferroelectric domain distribution. This effect was observed in our experiments as a hysteretic behaviour of electrical conductivity, $G_{sd}$, measured on devices made from twisted MoS$_2$ heterostructures with ferroelectric interfaces. To this end, we have prepared and studied several heterostructures consisting of marginally twisted pairs of mono-, bi- and tri-layers (L) of MoS2. Parallel ($\theta \approx 0°$) alignment for twisted 1L and 3L devices produces ferroelectric 3R interfaces between the twisted layers, however for 2L this alignment leads to predominantly centrosymmetric 2H stacking which has no electric polarisation. These heterostructures were encapsulated in hBN and double gated (the top gate was deposited only above the main channel, whereas the bottom gate (Si wafer) was global). At zero gate voltages, all the devices were found to be insulating, and required an applied electric field of ~0.1 V nm$^{-1}$ or, equivalently, the electron density of ~2×10$^{12}$ cm$^{-2}$ to start inducing mobile carriers. Therefore, to study the devices' conductivity $G_{sd}$, we applied a finite positive bottom gate voltage $V_b$, which was essential to induce electrical conductance in the contact regions, and then swept the top gate voltage $V_t$. Because of the generally low carrier mobility of MoS$_2$ channels at room temperature, only two-probe measurements were possible, with source-drain resistances reaching above MΩ in magnitude. Figs. 3f,g shows the hysteresis observed for twisted monolayer-on-monolayer and trilayer-on-trilayer MoS$_2$ (twisted at nominally zero degree in both cases). For further examples of hysteretic behaviour, see Supplementary Section 6. Sweeping $V_t$ controls both carrier concentration and the out-of-plane electric field which leads to redistribution between domains with opposite polarities, depending on whether positive or negative gate voltage is applied, as observed in our microscopy measurements in Fig. 1. Pinning of ferroelectric domain boundaries for up and down sweeps results in memory effects and hysteresis. As a result, the sample-averaged carrier densities for the same $V_t$ for up and down sweeps differ, leading to the observed difference in conductivities seen in Figs. 3f,g. Our reference devices made from bilayer-on-bilayer $MoS_2$ with predominant 2H stacking (Fig.3h) and exfoliated trilayer of 2H-$MoS_2$ (Supplementary Fig.S12b) showed a very similar electrical response to applied gate voltages but no discernible hysteresis. This is expected because the latter devices do not feature switchable ferroelectric interfaces, unlike the marginally twisted mono- and tri- layer heterostructures in Fig.3f,g. We also tested few-layer $MoS_2$ devices with 3R stacking. The material is an intrinsic ferroelectric, but no switching or hysteresis was observed for the prepared devices for all accessible gate voltages (Supplementary Fig.S12c). This is attributed to much stronger pinning at the 3R device edges than for movements of DW in marginally twisted $MoS_2$.

To summarise, our study shows that switchable ferroelectric behaviour is a generic property of heterostructures assembled from atomically thin TMDs with small twist angles providing a 3R interface. Having demonstrated this by studying field-driven domain evolution in both structural (BSECCI) and electronic (KPFM and transport) properties, we have also quantified the amount of charge transfer in the ferroelectric double layer at the interface and found good agreement with our theoretical modelling. These observations pave a way towards atomically thin electronic devices with memory effects and open up possibilities for the design of novel (opto)electronic devices. For example, strong light-matter coupling in TMDs[29] and the single-photon emission by defects in individual TMD layers[30] may offer a switchable single-photon emission capability.

## Acknowledgements:

We acknowledge support from EPSRC grants EP/V007033/1, EP/P009050/1, EP/S019367/1, EP/S030719/1, EP/V036343/1, the CDT Graphene-NOWNANO. In addition, we acknowledge support from European Union's Horizon 2020 Research and Innovation program: European Graphene Flagship Core3 Project (grant agreement 881603), European Quantum Flagship Project 2DSIPC (820378), ERC Synergy Grant Hetero2D, ERC Starter grant EvoluTEM (715502), ERC Consolidator grant QTWIST (101001515), Royal Society and Lloyd Register Foundation Nanotechnology grant. We also want to acknowledge the support of the UK government department for Business, Energy and Industrial Strategy through the UK national quantum technologies programme.

## Data Availability

Additional data related to this paper is available from the corresponding authors upon reasonable request.

## Code Availability

The computer code used for the image filtering is available from the corresponding authors upon reasonable request.

## Additional Information

Supplementary information is available in the online version of the paper. Reprints and permission information is available online at www.nature.com/reprints. Correspondence and requests for materials should be addressed to R.G.

## Competing Interest Statement

Authors declare no competing financial or non-financial interests.

## References:

(1) Martin, L. W.; Rappe, A. M. Thin-Film Ferroelectric Materials and Their Applications. *Nature Reviews Materials*. 2016, p 16087. https://doi.org/10.1038/natrevmats.2016.87.

(2) Mikolajick, T.; Slesazeck, S.; Mulaosmanovic, H.; Park, M. H.; Fichtner, S.; Lomenzo, P. D.; Hoffmann, M.; Schroeder, U. Next Generation Ferroelectric Materials for Semiconductor Process Integration and Their Applications. *J. Appl. Phys* **2021**, *129*, 100901. https://doi.org/10.1063/5.0037617.

(3) Bertolazzi, S.; Bondavalli, P.; Roche, S.; San, T.; Choi, S. Y.; Colombo, L.; Bonaccorso, F.; Samorì, P. Nonvolatile Memories Based on Graphene and Related 2D Materials. *Advanced Materials*. John Wiley & Sons, Ltd March 1, 2019, p 1806663. https://doi.org/10.1002/adma.201806663.

(4) Chang, K.; Liu, J.; Lin, H.; Wang, N.; Zhao, K.; Zhang, A.; Jin, F.; Zhong, Y.; Hu, X.; Duan, W.; et al. Discovery of Robust In-Plane Ferroelectricity in Atomic-Thick SnTe. *Science (80-. ).* **2016**, *353* (6296), 274–278. https://doi.org/10.1126/science.aad8609.

(5) Belianinov, A.; He, Q.; Dziaugys, A.; Maksymovych, P.; Eliseev, E.; Borisevich, A.; Morozovska, A.; Banys, J.; Vysochanskii, Y.; Kalinin, S. V. CuInP2S6 Room Temperature Layered Ferroelectric. *Nano Lett.* **2015**, *15* (6), 3808–3814. https://doi.org/10.1021/acs.nanolett.5b00491.

(6) Zhou, Y.; Wu, D.; Zhu, Y.; Cho, Y.; He, Q.; Yang, X.; Herrera, K.; Chu, Z.; Han, Y.; Downer, M. C.; et al. Out-of-Plane Piezoelectricity and Ferroelectricity in Layered α-In2Se3 Nanoflakes. *Nano Lett.* **2017**, *17* (9), 5508–5513. https://doi.org/10.1021/acs.nanolett.7b02198.

(7) Zheng, C.; Yu, L.; Zhu, L.; Collins, J. L.; Kim, D.; Lou, Y.; Xu, C.; Li, M.; Wei, Z.; Zhang, Y.; et al. Room Temperature In-Plane Ferroelectricity in van Der Waals In2Se3. *Sci. Adv.* **2018**, *4* (7), eaar7720. https://doi.org/10.1126/sciadv.aar7720.

(8) Li, Y.; Gong, M.; Zeng, H. Atomically Thin In2Se3: An Emergent Two-Dimensional Room Temperature Ferroelectric Semiconductor. *Journal of Semiconductors*. 2019. https://doi.org/10.1088/1674-4926/40/6/061002.

(9) Woods, C. R.; Ares, P.; Nevison-Andrews, H.; Holwill, M. J.; Fabregas, R.; Guinea, F.; Geim, A. K.; Novoselov, K. S.; Walet, N. R.; Fumagalli, L. Charge-Polarized Interfacial Superlattices in Marginally Twisted Hexagonal Boron Nitride. *Nat. Commun.* **2021**, *12* (1), 1–7. https://doi.org/10.1038/s41467-020-20667-2.

(10) Yasuda, K.; Wang, X.; Watanabe, K.; Taniguchi, T.; Jarillo-Herrero, P. Stacking-Engineered Ferroelectricity in Bilayer Boron Nitride. *Science (80-. ).* **2021**, *372* (6549), eabd3230. https://doi.org/10.1126/science.abd3230.

(11) Vizner Stern, M.; Waschitz, Y.; Cao, W.; Nevo, I.; Watanabe, K.; Taniguchi, T.; Sela, E.; Urbakh, M.; Hod, O.; Ben Shalom, M. Interfacial Ferroelectricity by van Der Waals Sliding. *Science (80-. ).* **2021**, *372* (6549), eabe8177. https://doi.org/10.1126/science.abe8177.

(12) de la Barrera, S. C.; Cao, Q.; Gao, Y.; Gao, Y.; Bheemarasetty, V. S.; Yan, J.; Mandrus, D. G.; Zhu, W.; Xiao, D.; Hunt, B. M. Direct Measurement of Ferroelectric Polarization in a Tunable Semimetal. **2020**, 1–14.

(13) Ferreira, F.; Enaldiev, V. V.; Fal'ko, V. I.; Magorrian, S. J.; Fal'ko, V. I.; Magorrian, S. J. Weak Ferroelectric Charge Transfer in Layer-Asymmetric Bilayers of 2D Semiconductors. *Sci. Rep.* **2021**, *11* (1), 1–10. https://doi.org/10.1038/s41598-021-92710-1.

(14) Weston, A.; Zou, Y.; Enaldiev, V.; Summerfield, A.; Clark, N.; Zólyomi, V.; Graham, A.; Yelgel, C.; Magorrian, S.; Zhou, M.; et al. Atomic Reconstruction in Twisted Bilayers of Transition Metal Dichalcogenides. *Nat. Nanotechnol.* **2020**, *15* (7), 592–597. https://doi.org/10.1038/s41565-020-0682-9.

(15) Sung, J.; Zhou, Y.; Scuri, G.; Zólyomi, V.; Andersen, T. I.; Yoo, H.; Wild, D. S.; Joe, A. Y.; Gelly, R. J.; Heo, H.; et al. Broken Mirror Symmetry in Excitonic Response of Reconstructed Domains in Twisted MoSe2/MoSe2 Bilayers. *Nat. Nanotechnol.* **2020**. https://doi.org/10.1038/s41565-020-0728-z.

(16) Wang, G.; Chernikov, A.; Glazov, M. M.; Heinz, T. F.; Marie, X.; Amand, T.; Urbaszek, B. Excitons in Atomically Thin Transition Metal Dichalcogenides. *Rev. Mod. Phys.* **2017**, *90* (2), 021001. https://doi.org/10.1103/RevModPhys.90.021001.

(17) Rooney, A. P.; Kozikov, A.; Rudenko, A. N.; Prestat, E.; Hamer, M. J.; Withers, F.; Cao, Y.; Novoselov, K. S.; Katsnelson, M. I.; Gorbachev, R.; et al. Observing Imperfection in Atomic Interfaces for van Der Waals Heterostructures. *Nano Lett.* **2017**, *17* (9), 5222–5228. https://doi.org/10.1021/acs.nanolett.7b01248.

(18) Weston, A.; Zou, Y.; Enaldiev, V.; Summerfield, A.; Clark, N.; Zólyomi, V.; Graham, A.; Yelgel, C.; Magorrian, S.; Zhou, M.; et al. Atomic Reconstruction in Twisted Bilayers of Transition Metal Dichalcogenides. *Nat. Nanotechnol.* **2020**, *15* (7), 592–597. https://doi.org/10.1038/s41565-020-0682-9.

(19) Rosenberger, M. R.; Chuang, H.-J.; Phillips, M.; Oleshko, V. P.; McCreary, K. M.; Sivaram, S. V.; Hellberg, C. S.; Jonker, B. T. Twist Angle-Dependent Atomic Reconstruction and Moiré Patterns in Transition Metal Dichalcogenide Heterostructures. *ACS Nano* **2020**, *14* (4), 4550–4558. https://doi.org/10.1021/acsnano.0c00088.

(20) Kim, K.; Yankowitz, M.; Fallahazad, B.; Kang, S.; Movva, H. C. P.; Huang, S.; Larentis, S.; Corbet, C. M.; Taniguchi, T.; Watanabe, K.; et al. Van Der Waals Heterostructures with High Accuracy Rotational Alignment. *Nano Lett.* **2016**, *16* (3), 1989–1995. https://doi.org/10.1021/acs.nanolett.5b05263.

(21) Uri, A.; Grover, S.; Cao, Y.; Crosse, J. A.; Bagani, K.; Rodan-Legrain, D.; Myasoedov, Y.; Watanabe, K.; Taniguchi, T.; Moon, P.; et al. Mapping the Twist-Angle Disorder and Landau Levels in Magic-Angle Graphene. *Nature* **2020**, *581* (7806), 47–52. https://doi.org/10.1038/s41586-020-2255-3.

(22) Wilkinson, A. J.; Hirsch, P. B. Electron Diffraction Based Techniques in Scanning Electron Microscopy of Bulk Materials. *Micron* **1997**, *28* (4), 279–308. https://doi.org/10.1016/S0968-4328(97)00032-2.

(23) Andersen, T. I.; Scuri, G.; Sushko, A.; De Greve, K.; Sung, J.; Zhou, Y.; Wild, D. S.; Gelly,

R. J.; Heo, H.; Bérubé, D.; et al. Excitons in a Reconstructed Moiré Potential in Twisted WSe2/WSe2 Homobilayers. *Nat. Mater.* **2021**, *20* (4), 480–487. https://doi.org/10.1038/s41563-020-00873-5.

(24) Giannozzi, P.; Baseggio, O.; Bonfà, P.; Brunato, D.; Car, R.; Carnimeo, I.; Cavazzoni, C.; De Gironcoli, S.; Delugas, P.; Ferrari Ruffino, F.; et al. Quantum ESPRESSO toward the Exascale. *J. Chem. Phys.* **2020**, *152* (15), 154105. https://doi.org/10.1063/5.0005082.

(25) Ahmed, F.; Heo, S.; Yang, Z.; Ali, F.; Ra, C. H.; Lee, H. I.; Taniguchi, T.; Hone, J.; Lee, B. H.; Yoo, W. J. Dielectric Dispersion and High Field Response of Multilayer Hexagonal Boron Nitride. *Adv. Funct. Mater.* **2018**, *28* (40), 1804235. https://doi.org/10.1002/adfm.201804235.

(26) Laturia, A.; Van de Put, M. L.; Vandenberghe, W. G. Dielectric Properties of Hexagonal Boron Nitride and Transition Metal Dichalcogenides: From Monolayer to Bulk. *npj 2D Mater. Appl.* **2018**, *2* (1), 1–7. https://doi.org/10.1038/s41699-018-0050-x.

(27) Enaldiev, V. V.; Zólyomi, V.; Yelgel, C.; Magorrian, S. J.; Fal'ko, V. I. Stacking Domains and Dislocation Networks in Marginally Twisted Bilayers of Transition Metal Dichalcogenides. *Phys. Rev. Lett.* **2020**, *124* (20), 206101. https://doi.org/10.1103/PhysRevLett.124.206101.

(28) Ferreira, F.; Magorrian, S. J.; Enaldiev, V. V; Ruiz-Tijerina, D. A.; Fal'ko, V. I. Band Energy Landscapes in Twisted Homobilayers of Transition Metal Dichalcogenides. *Appl. Phys. Lett.* **2021**, *118* (24). https://doi.org/10.1063/5.0048884.

(29) Britnell, L.; Ribeiro, R. M. M.; Eckmann, A.; Jalil, R.; Belle, B. D. D.; Mishchenko, A.; Kim, Y.-J. J.; Gorbachev, R. V. V; Georgiou, T.; Morozov, S. V. V; et al. Strong Light-Matter Interactions in Heterostructures of Atomically Thin Films. *Science (80-. ).* **2013**, *340* (6138), 1311–1314. https://doi.org/10.1126/science.1235547.

(30) Koperski, M.; Nogajewski, K.; Arora, A.; Cherkez, V.; Mallet, P.; Veuillen, J.-Y.; Marcus, J.; Kossacki, P.; Potemski, M. Single Photon Emitters in Exfoliated WSe$_2$ Structures. *Nat. Nanotechnol.* **2015**, *10* (6), 503–506. https://doi.org/10.1038/nnano.2015.67.